\documentclass[twoside,slac_one]{revtex4}
\usepackage{graphicx}
\usepackage{fancyhdr}
\usepackage{amsmath} 
\usepackage{bm}
\usepackage{amsxtra}
\usepackage{amssymb}
\usepackage{amsthm}
\usepackage{latexsym}
\usepackage{lscape}

\begin{document}

\title{Single top quark production at D0}

%

\author{R. Schwienhorst\footnote{For the D0 Collaboration} }
\affiliation{Department of Physics and Astronomy, 
Michigan State University, East Lansing, USA}

\begin{abstract}
Updates of electroweak single top quark production measurements
by the D0 collaboration are presented using 5.4~fb$^{-1}$ of proton-antiproton 
collision data from the Tevatron at Fermilab. Measurements of the $t$-channel,
$s$-channel and combined single top quark production cross section are presented,
including an updated lower limit on the CKM matrix element $|V_{tb}|$.
Also reported are results from searches for gluon-quark flavor-changing neutral 
currents and $W'$~boson production.
\end{abstract}

\maketitle

\thispagestyle{fancy}


\section{Introduction}
\begin{figure}[!h!tbp]
\includegraphics[width=100mm]{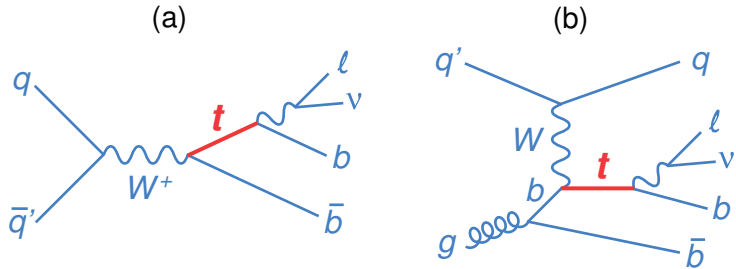}
\caption{Representative Feynman diagrams for (a) $s$-channel
single top quark production and (b) $t$-channel production, including
the top quark decay.}
\label{fig:feynman}
\end{figure}

The production of single top quark events via the electroweak interaction was 
reported in 2009 by the D0~\cite{Abazov:2009ii} and CDF~\cite{Aaltonen:2009jj} 
collaborations. A measurement of the single top quark production cross section 
provides a direct measurement of the the quark mixing matrix element 
$|V_{tb}|$~\cite{singletop-vtb-jikia}. It also serves as a probe of the $Wtb$ 
coupling~\cite{Chen:2005vr,dudko-boos,singletop-wtb-heinson,d0-singletop-wtb,Abazov:2009ky}
and is sensitive to several models of new physics~\cite{Tait:2000sh}.
Single top quark production proceeds via the $t$-channel 
exchange of a virtual $W$~boson between a light quark line and a heavy quark 
line ($tqb$) and the $s$-channel production and decay of a virtual $W$~boson 
($tb$), shown in Fig.~\ref{fig:feynman}.
~ 

Here we present an updated measurement of the combined single top quark
production cross section using 5.4~fb$^{-1}$ of data collected by the D0
experiment at the Tevatron proton-antiproton collider at 
Fermilab~\cite{Abazov:2011pt}. The $s$-channel and $t$-channel modes are 
measured separately and the first observation of $t$-channel production
is reported~\cite{Abazov:2011rz}.

\section{SM single top quark measurement}

\subsection{Event selection}
The single top quark final state consists of a lepton (electron or muon)
and a neutrino from the $W$~boson decay, a $b$~quark from the top quark decay 
and an additional light quark ($t$-channel) or $b$~quark ($s$-channel). 
The event selection for all analysis channels requires large 
$\slash\kern-.7emE_T$ and two to four jets, at least one of which is $b$-tagged. 

The backgrounds to this signature are dominated by $W$~bosons produced in 
association with jets ($W$+jets), with smaller contributions from $t\overline{t}$
pairs. Multijet events also contribute to the background when a jet is 
misidentified as an isolated electron or a heavy-flavor quark decay results in
an isolated lepton. Diboson ($WW$, $WZ$, $ZZ$) and $Z$+jets contribute smaller
backgrounds. In total,
about 8,000 events are selected with an expected SM signal of about 400 events.

\begin{figure}[!h!tbp]
\centering
\includegraphics[width=60mm]{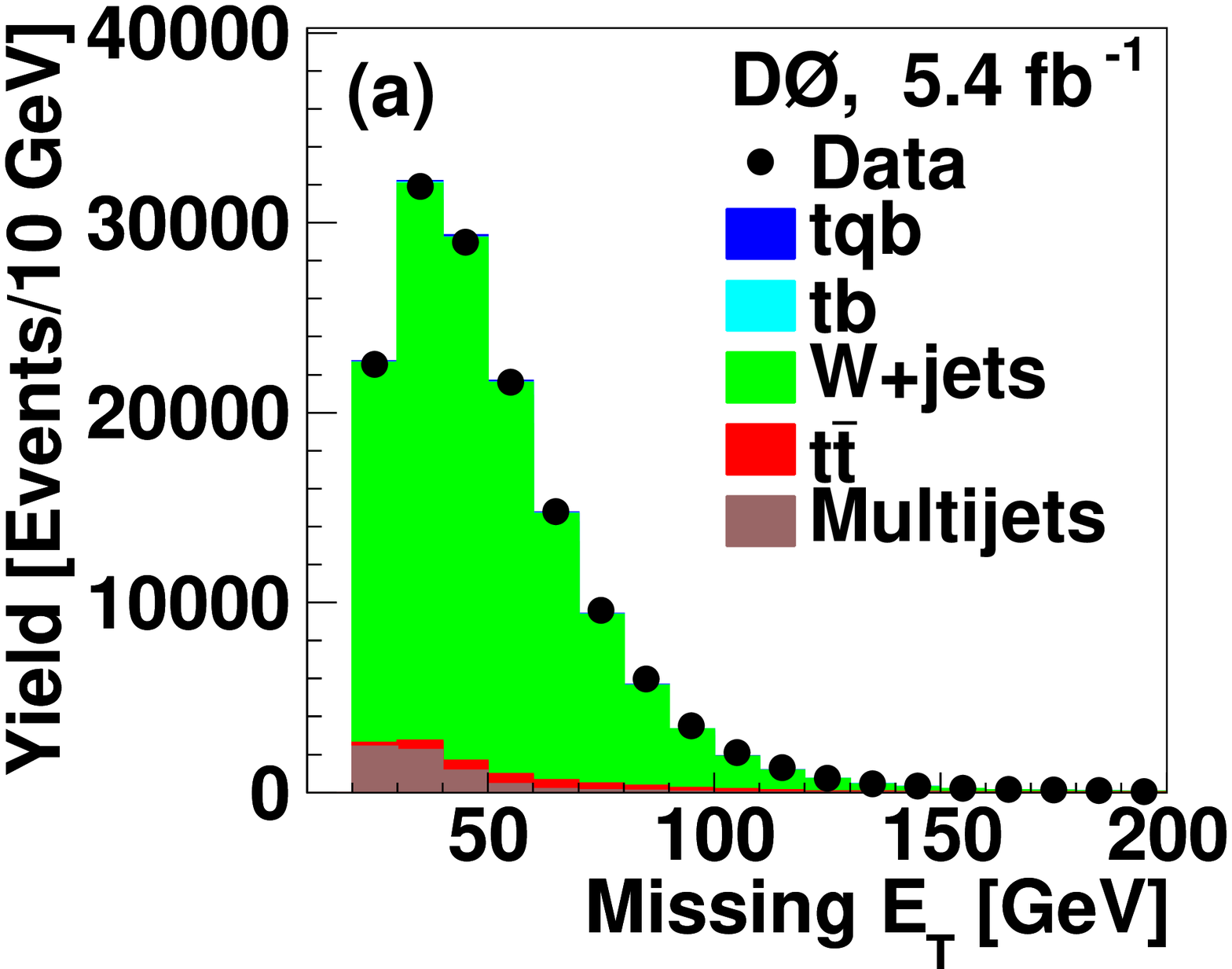}
\includegraphics[width=60mm]{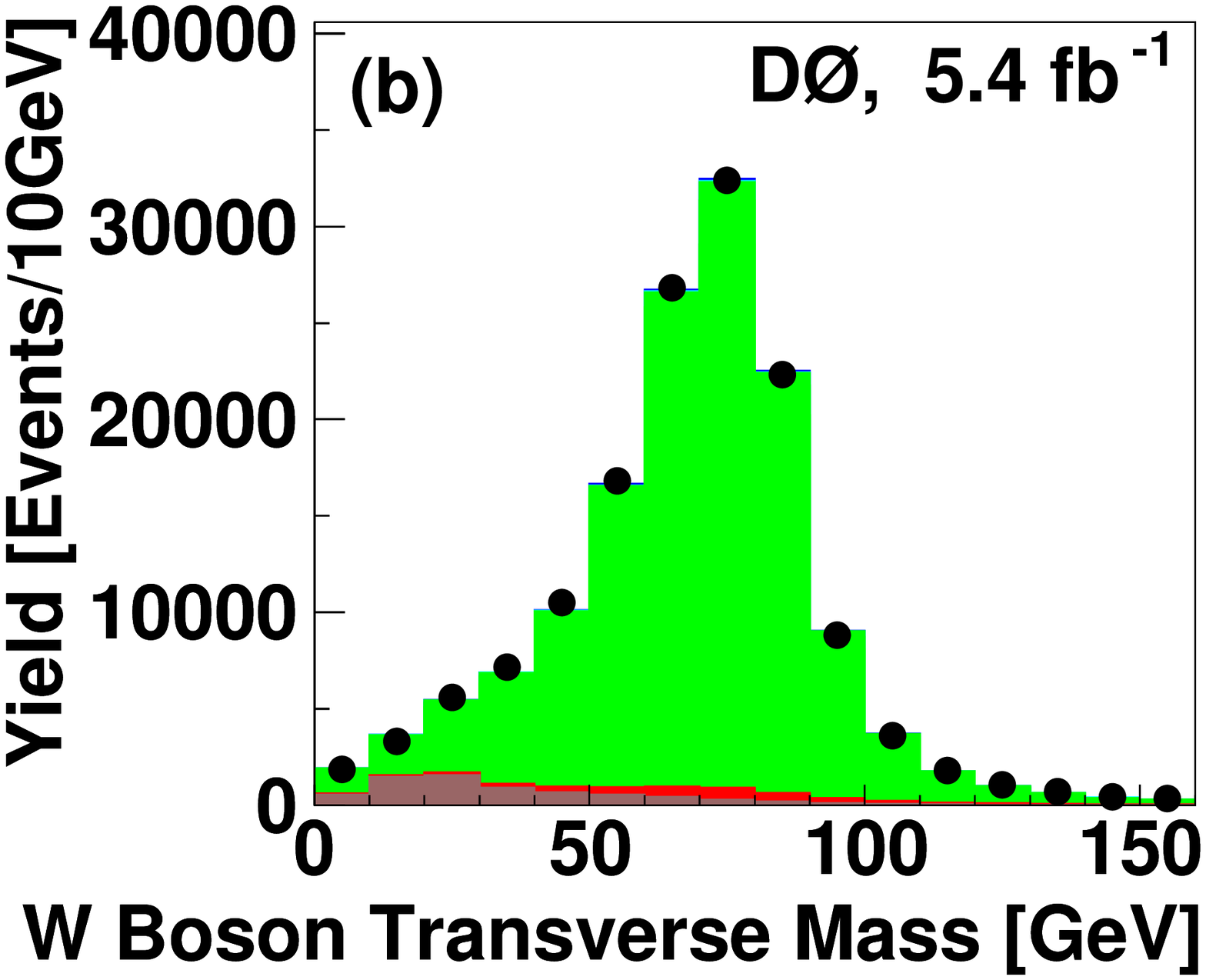}
\includegraphics[width=60mm]{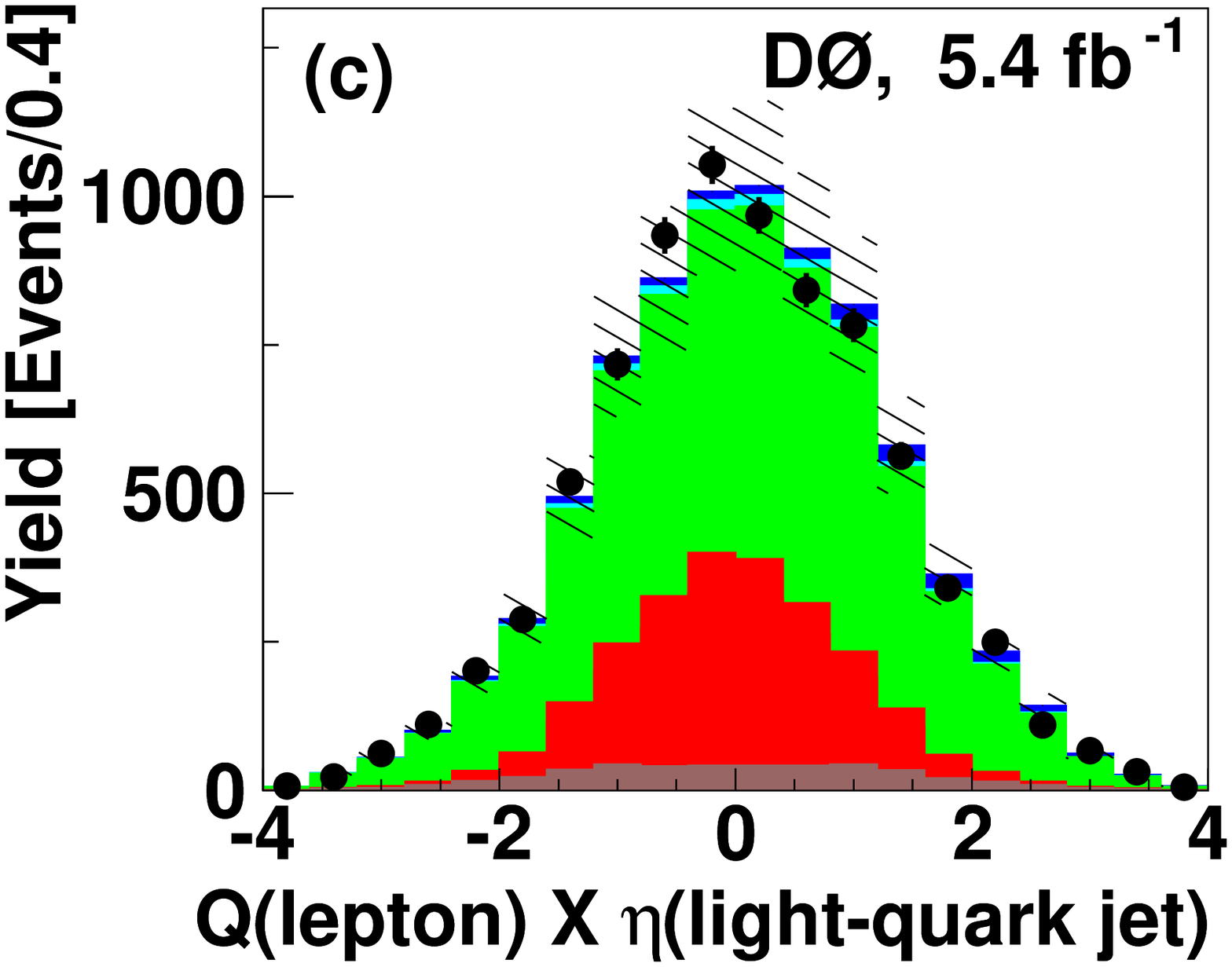}
\includegraphics[width=60mm]{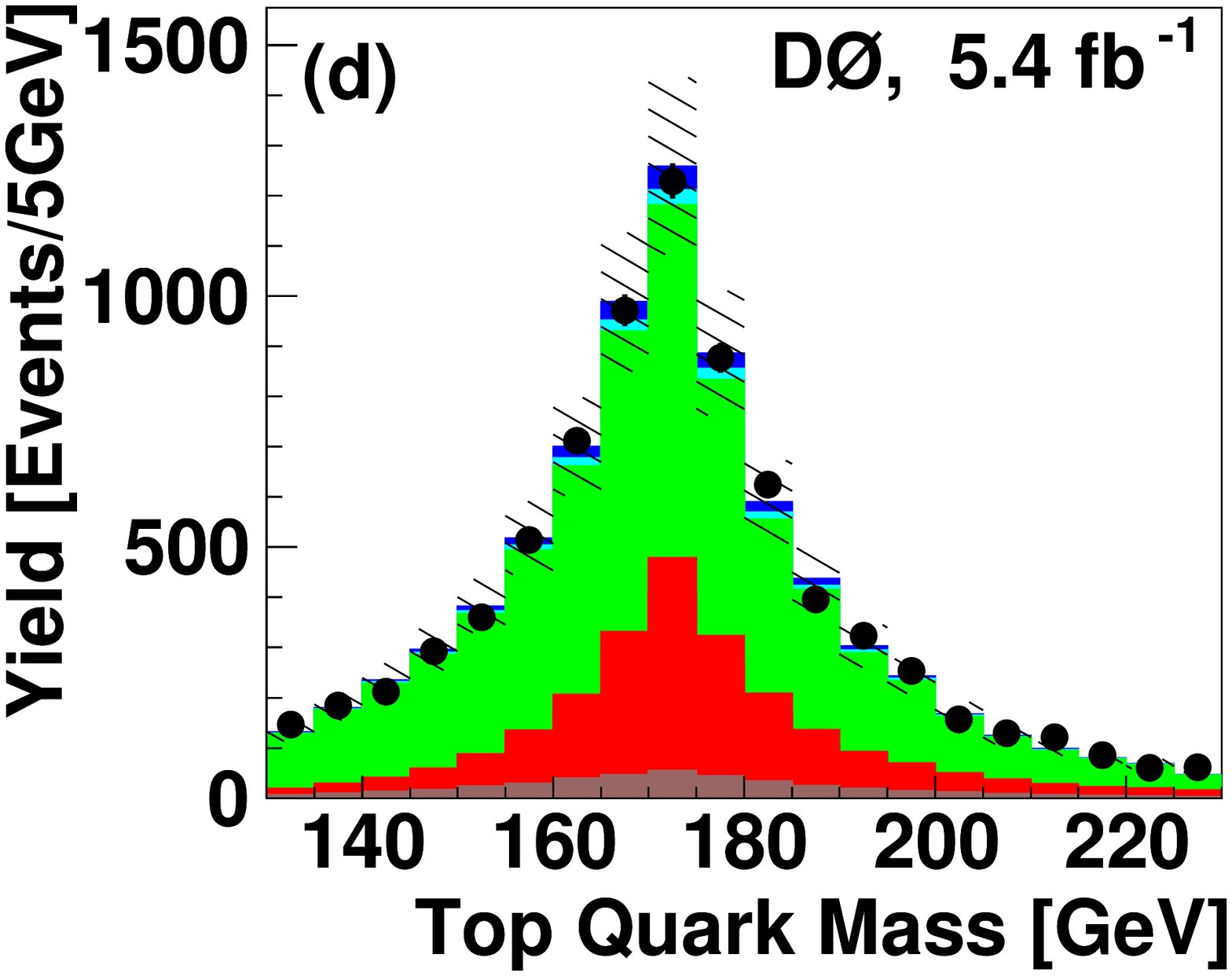}
\includegraphics[width=60mm]{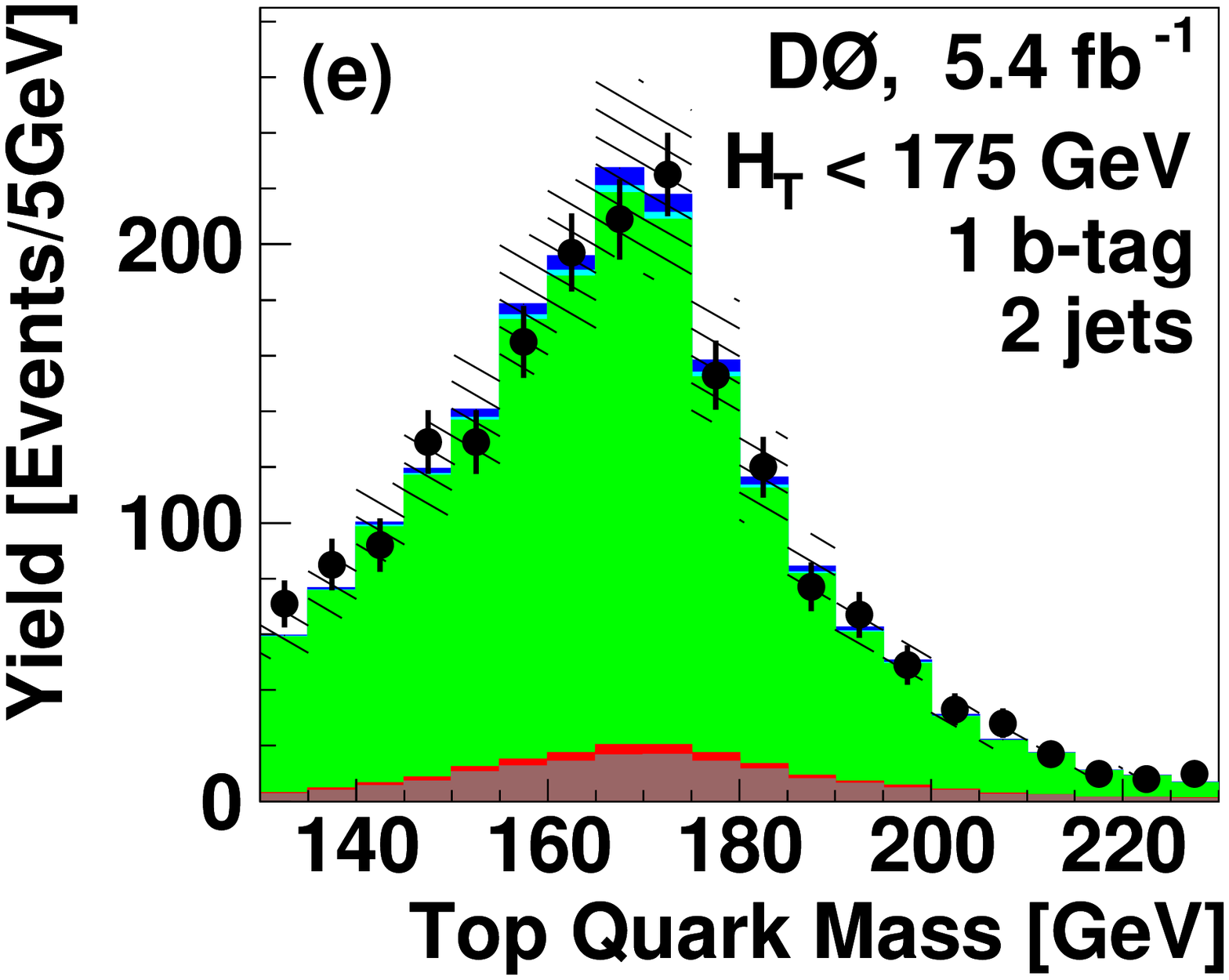}
\includegraphics[width=60mm]{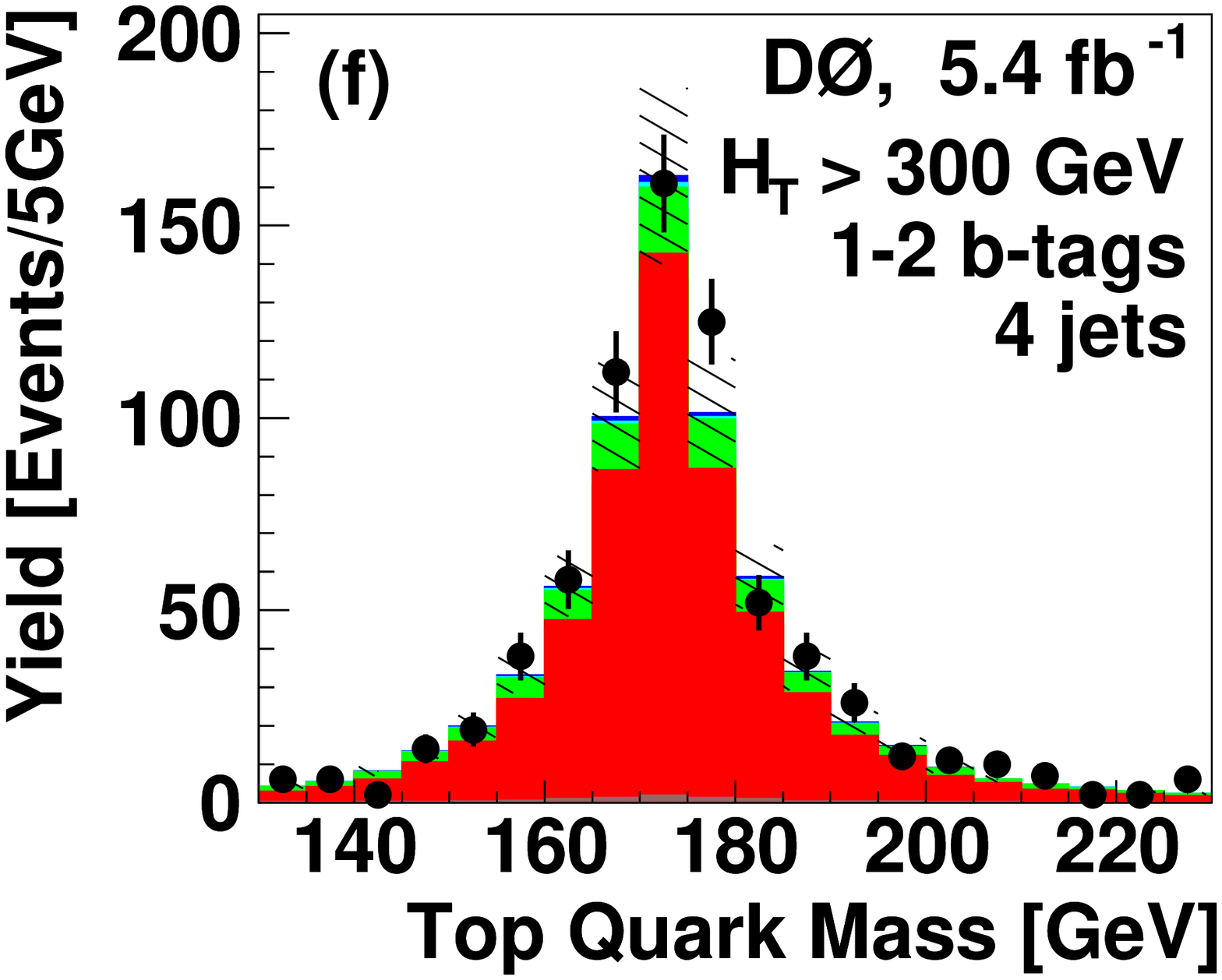}
\caption{
Comparisons between the data and the background model for 
(a) $\slash\kern-.7emE_T$, 
(b) $W$~boson transverse mass, 
(c) light quark jet pseudorapidity multiplied by lepton charge,  
(d--f) reconstructed top quark mass.
Subfigures (a) and (b) are before $b$-tagging, (c) and (d) are after $b$-tagging, 
(e) is a control sample dominated by $W$+jets, 
and (f) is a control sample dominated by $t\overline{t}$ pair events. 
The hatched bands show the $\pm 1\sigma$ uncertainty on the background prediction
for  distributions obtained after $b$-jet identification (c--f). 
} 
\label{fig:variables}
\end{figure}

In addition to the single top analysis sample, several cross-check samples are
defined: a without any $b$-tag requirements; a sample containing exactly
two jets and sum of transverse momenta of all objects $H_T<175$~GeV; 
and a sample containing exactly four jets and requiring $H_T>300$~GeV.
Figure~\ref{fig:variables} shows a comparison between the data and the 
signal+background model for several discriminating variables used in the
analysis in the different samples.
~ 

\subsection{Multivariate analysis}
\begin{figure}
\centering
\includegraphics[width=60mm]{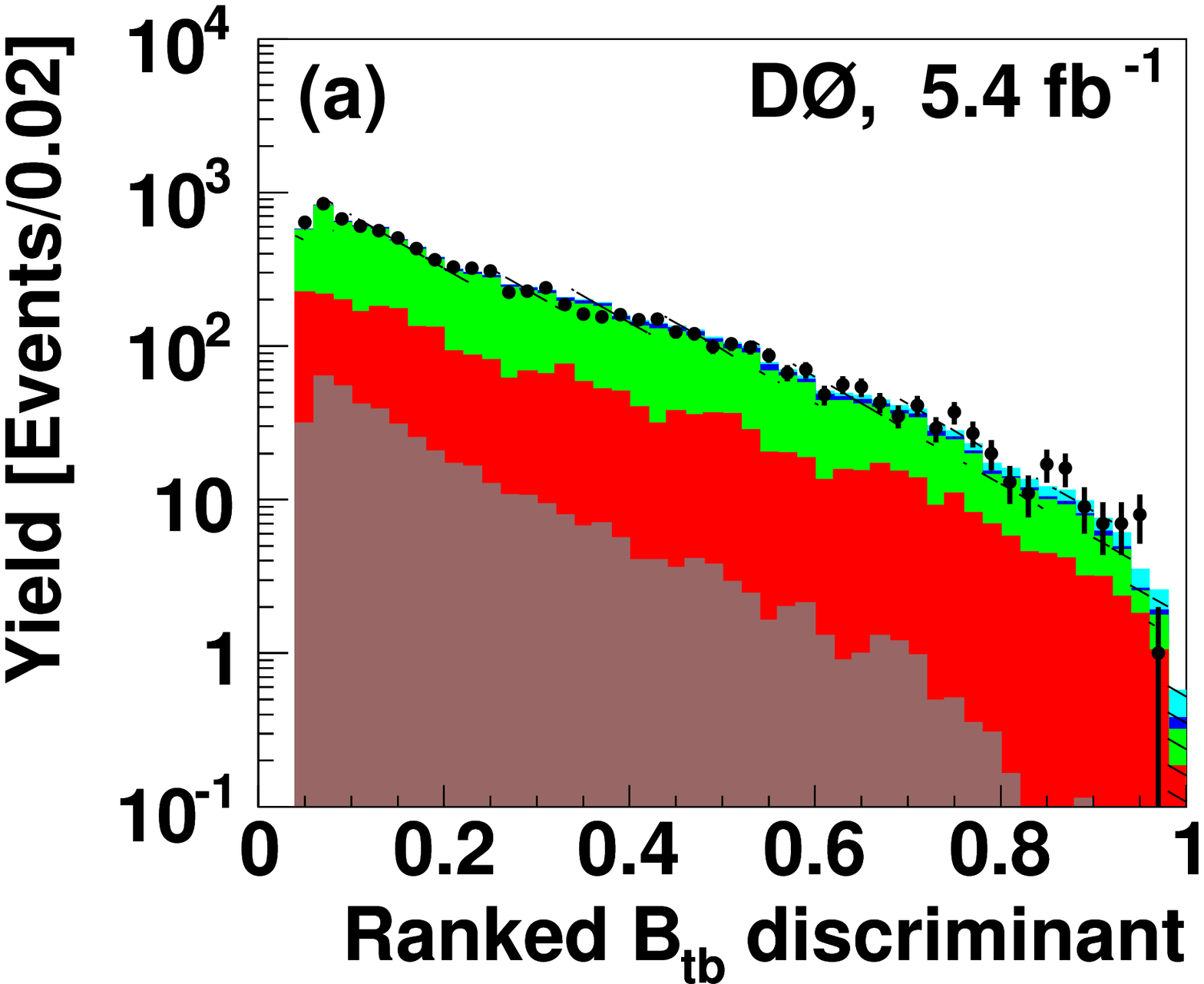}
\includegraphics[width=60mm]{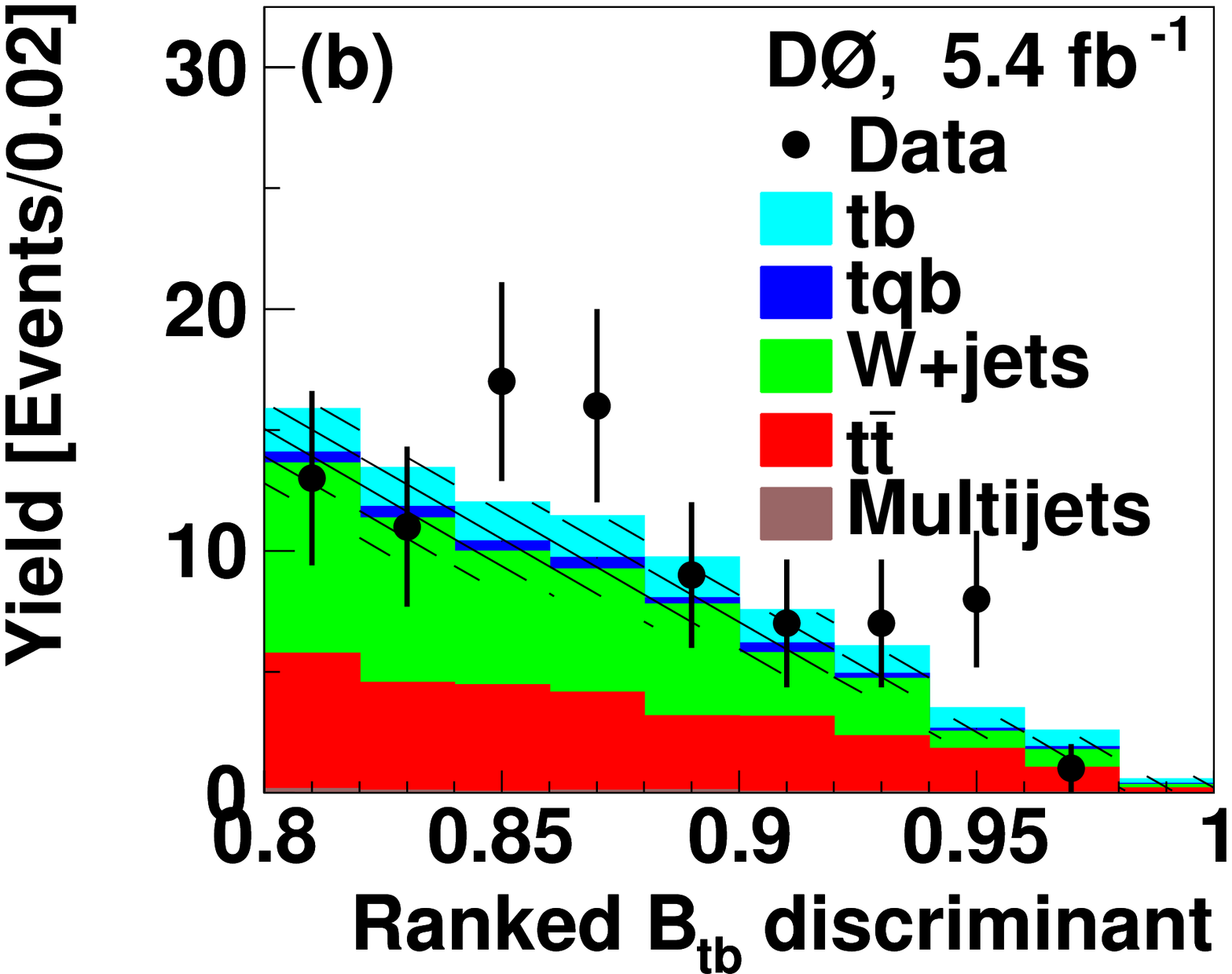}
\includegraphics[width=60mm]{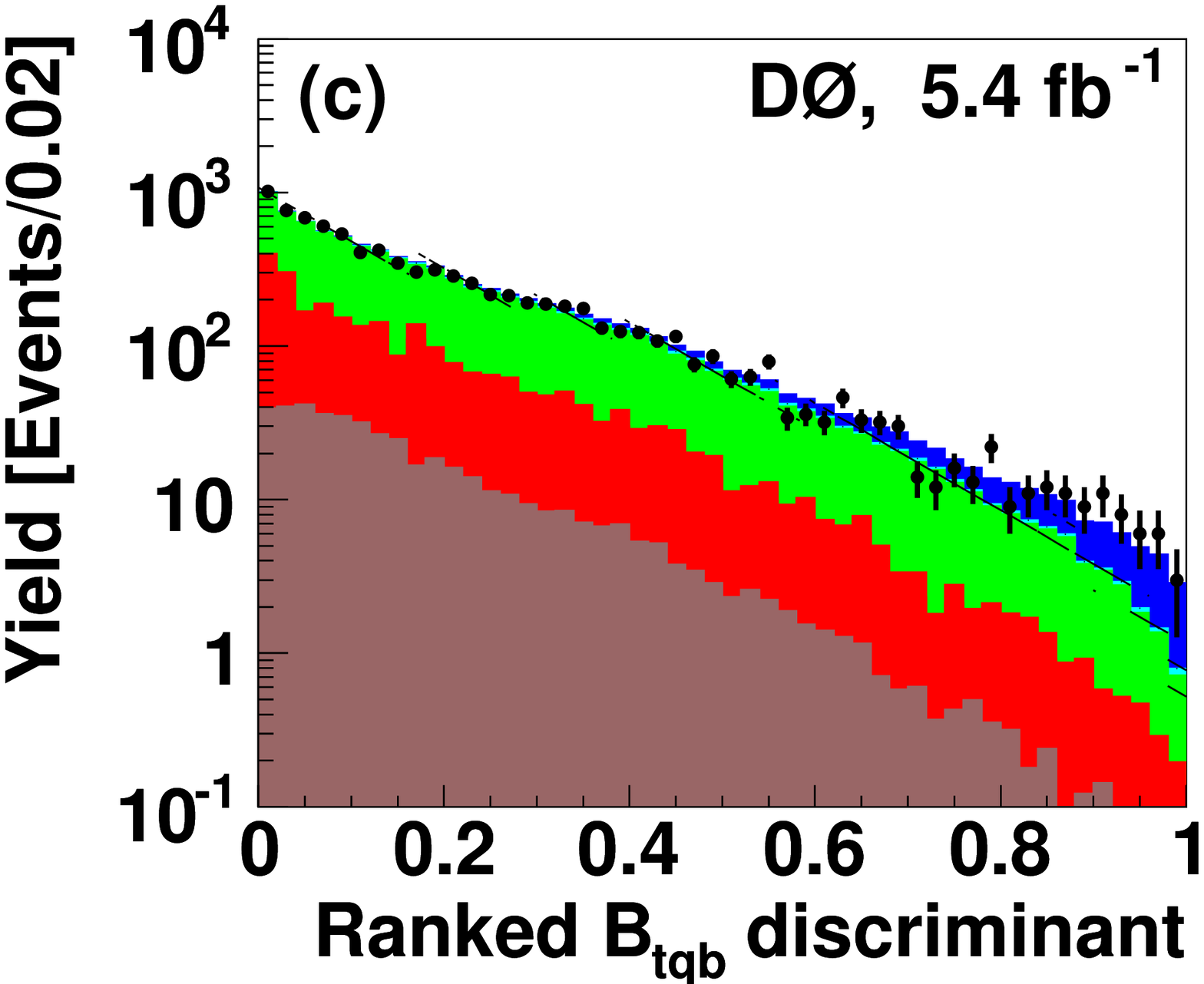}
\includegraphics[width=60mm]{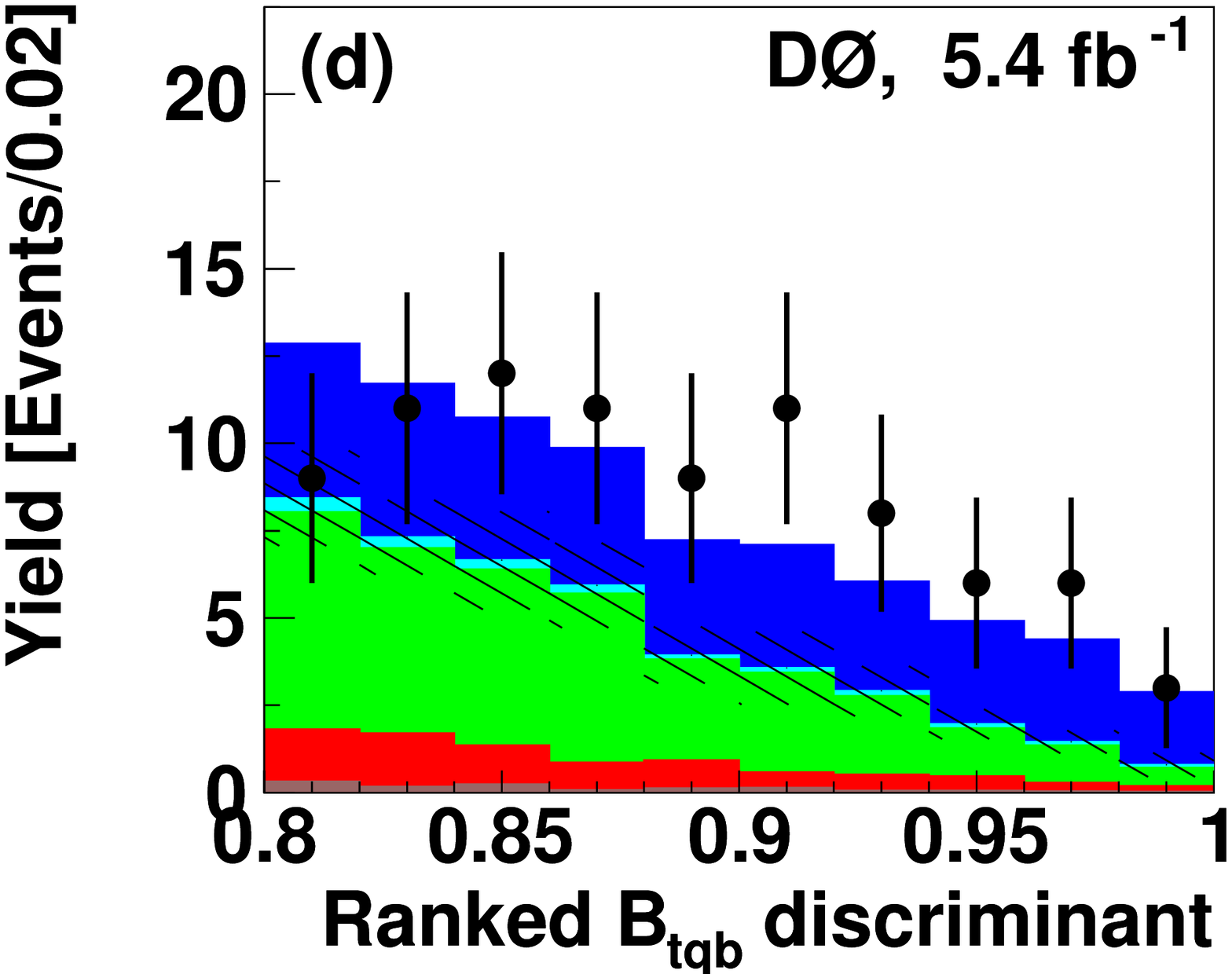}
\includegraphics[width=60mm]{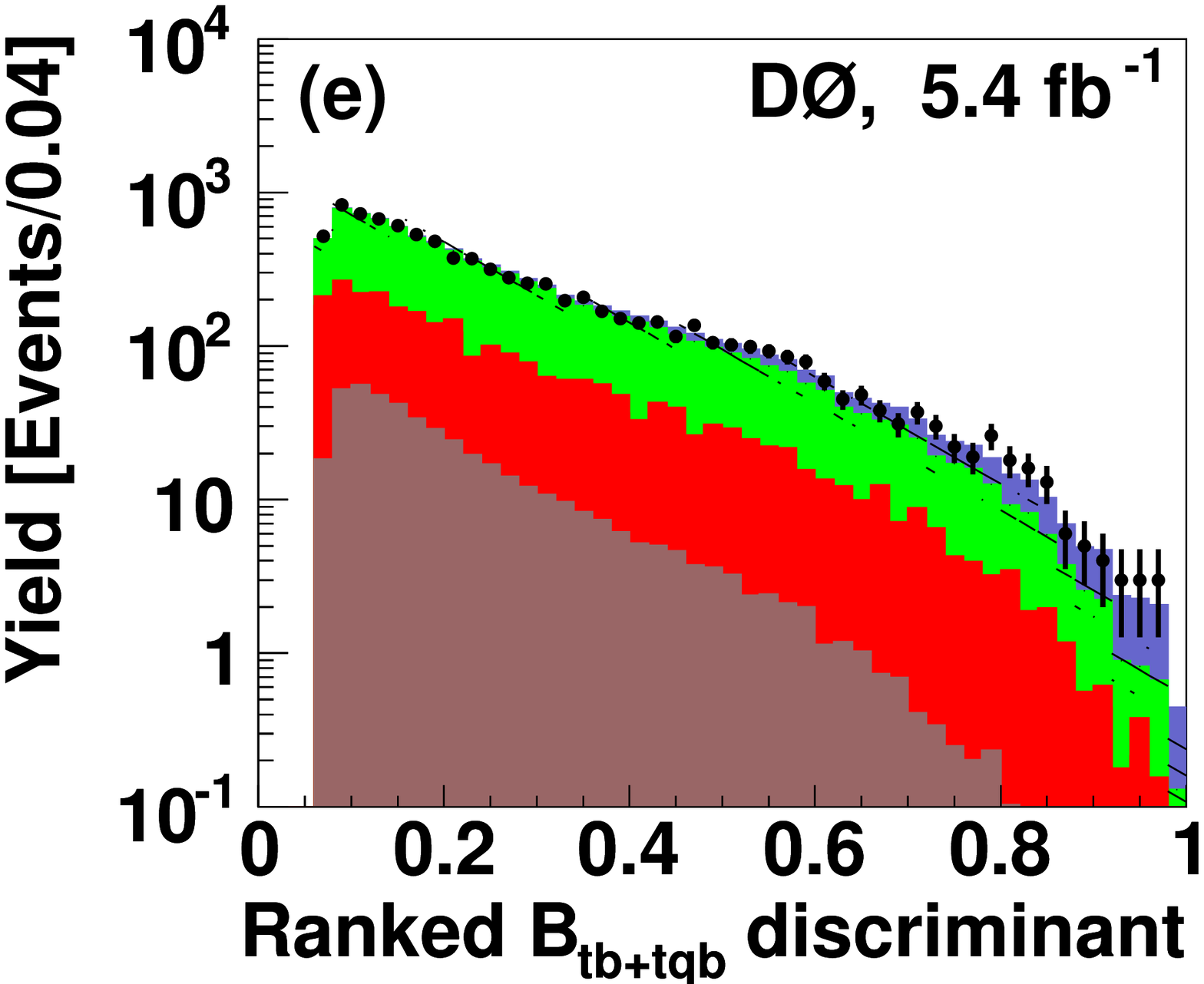}
\includegraphics[width=60mm]{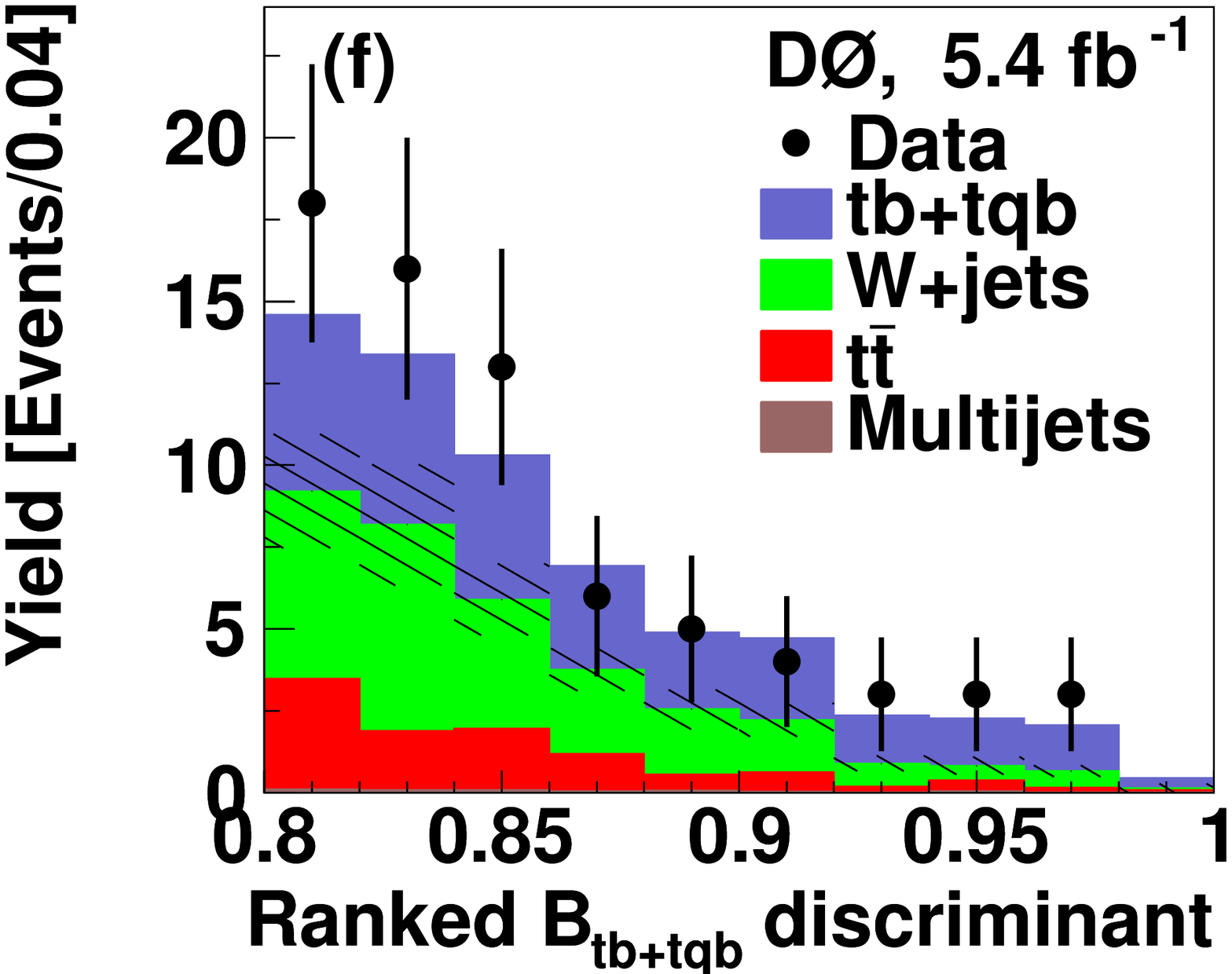}
\caption{Distributions of the combined multivariate discriminants for
(a, b) $s$-channel, 
(c, d) $t$-channel, 
and (e, f) combined single top production. 
The full distributions are shown in (a, c, e) and the signal region is shown
in (b, d, f).
The $tb$, $tqb$, and $tb$+$tqb$ contributions are normalized to their measured 
cross sections.
The hatched bands show the $\pm 1\sigma$ uncertainty on the background 
prediction.}
\label{fig:discriminants}
\end{figure}

The expected single top quark signal is small compared to the statistical data
uncertainty and the uncertainty on the background prediction. Thus, 
multivariate analysis methods are employed to extract the single top quark 
signal. Several different multivariate methods (Boosted decision trees,
Bayesian neural networks and a Neuro-evolution network) are trained and then 
combined into one filter. This training is done separately for the combined
single top analysis and for the dedicated $t$-channel and $s$-channel analyses.
The resulting discriminants are shown in Fig.~\ref{fig:discriminants}.
~

\subsection{Cross section measurement}
The cross section is measured using a Bayesian approach, including all systematic
uncertainties and their correlations~\cite{d0-prd-2008}. The expected and observed
cross section measurements are given in Table~\ref{tab:xsections}. The separate
$t$-channel and $s$-channel cross sections are extracted assuming the SM 
contribution for the $s$-channel and $t$-channel, respectively.

The $t$-channel signal is observed with a significance of more than five
standard deviations, and the combined single top measurement is the most
precise measurement to date.

\begin{table}[!h!tbp]
\begin{center}
\caption{Expected and observed cross sections in pb for $tb$, $tqb$, and 
$tb$+$tqb$ production. All results assume a top quark mass of $172.5\;\rm GeV$
and the expected cross section normalize the single top signal to its 
theoretical prediction~\cite{singletop-xsec-kidonakis}.}
\label{tab:xsections}
\begin{tabular}{|l|c|c|}
\hline
Discriminant & Expected      & Observed      \\
             & cross section & cross section \\
\hline
$tb$ & $1.12^{+0.45}_{-0.43}$ pb      & $0.68^{+0.38}_{-0.35}$ pb \\
\hline
$tqb$ & $2.43^{+0.67}_{-0.61}$ pb     & $2.86^{+0.69}_{-0.63}$ pb \\
\hline
$tb+tqb$  & $3.49^{+0.77}_{-0.71}$ pb & $3.43^{+0.73}_{-0.74}$ pb \\
\hline
\end{tabular}
\end{center}
\end{table}

\subsection{Vtb}
The single top quark production cross section is proportional to the CKM matrix 
element $|V_{tb}|^2$, allowing for a $|V_{tb}|$ measurement without having to 
assume three quark generations or CKM matrix unitarity. 
We only assume that SM sources for single top quark production and that top 
quarks decay exclusively to $Wb$, as well as that the $Wtb$ interaction is 
CP-conserving and of the $V-A$ type.
We set a lower limit at the 95\% confidence level of $|V_{tb}| > 0.79$.

\section{Searches for new physics}

\subsection{Search for flavor-changing neutral currents}
\begin{figure}
\centering
\includegraphics[width=60mm]{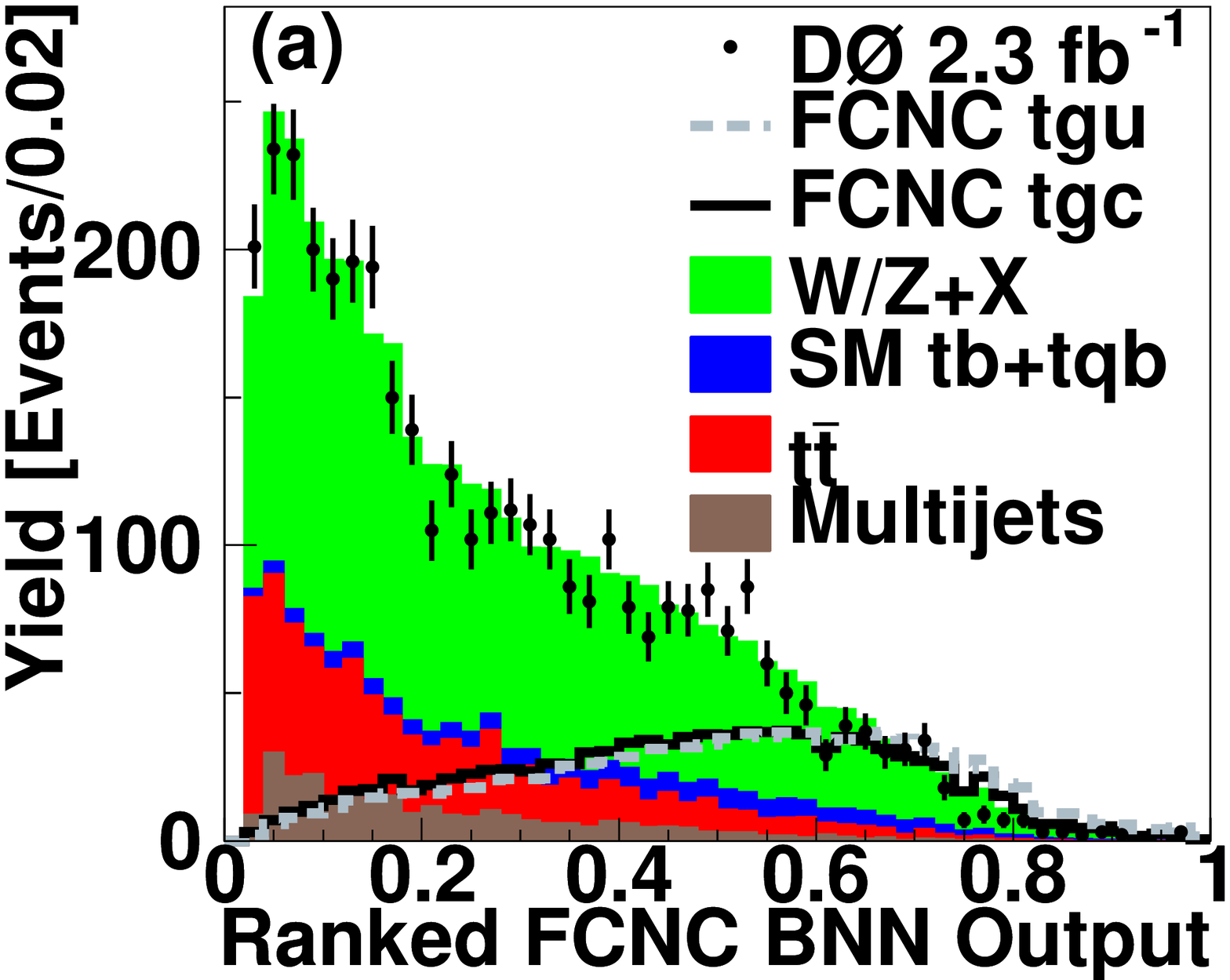}
\includegraphics[width=60mm]{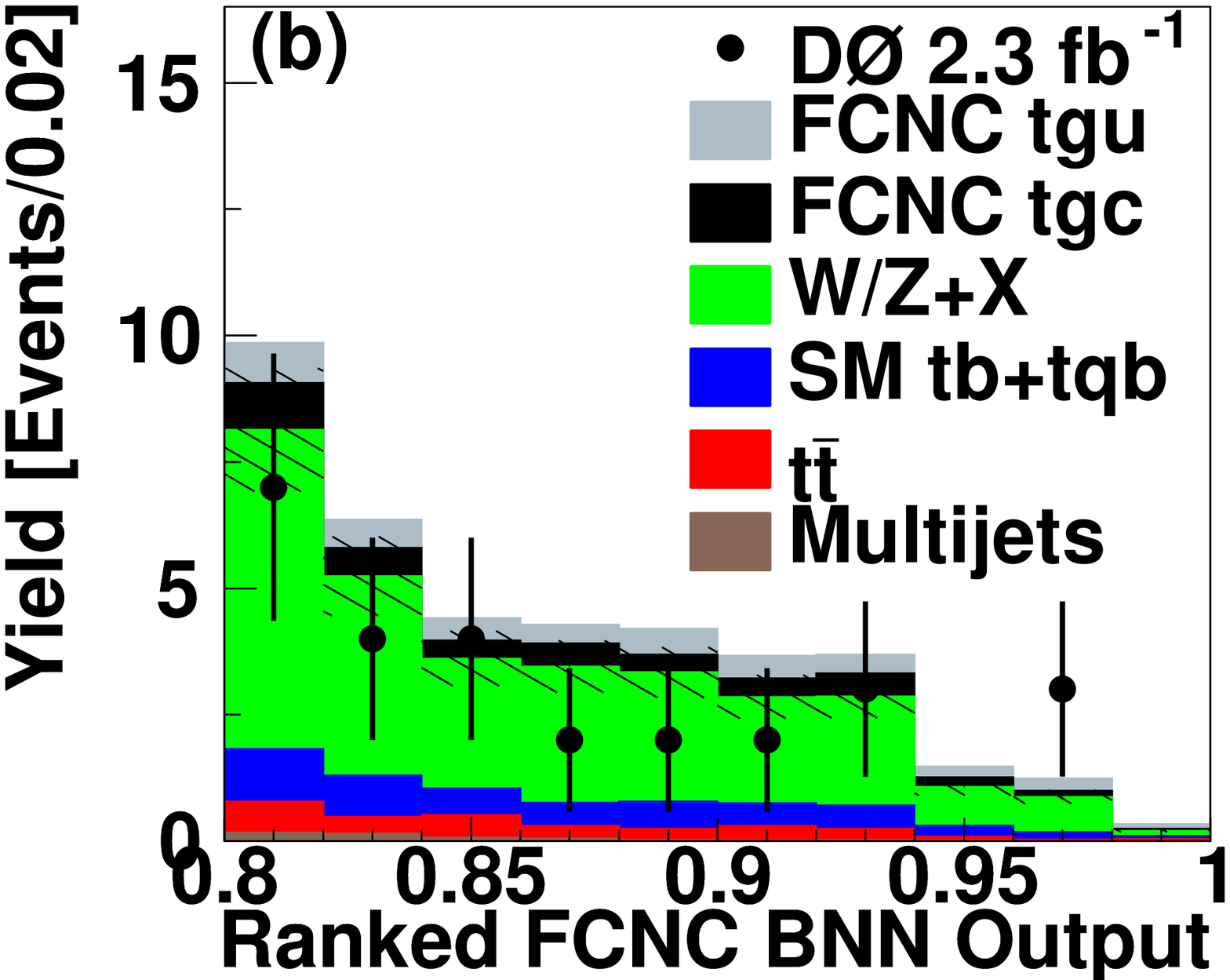}
\caption{Distributions of the FCNC discriminant for 
(a) the full discriminant range and (b) the signal region.
\label{fig:fcnc}}
\end{figure}

The $t$-channel final state is sensitive to flavor-changing neutral currents 
(FCNC) via quark-gluon couplings. In FCNC interactions a gluon vertex couples 
an up~quark ($tgu$) or a charm~quark ($tgc$) to the top quark. The rate for 
these events at the Tevatron is large because the initial state contains two 
light quarks. The D0~collaboration searched for FCNC interactions in 
2.3~fb$^{-1}$~\cite{Abazov:2010qk}, using the single top quark observation
analysis sample~\cite{Abazov:2009ii}. 
A Bayesian Neural Network (BNN) is used to 
separate the FCNC signal from the large backgrounds.
The BNN output distribution is shown in Fig.~\ref{fig:fcnc}.
The observed limits on the FCNC couplings are 
$\kappa_{tgu}/\Lambda < 0.013$~TeV$^{-1}$ and
$\kappa_{tgc}/\Lambda < 0.057$~TeV$^{-1}$, 
without making assumptions about the $tgc$ and $tgu$ couplings,
respectively. These can be translated into limits on top quark decay branching 
fractions, which are 
$\mathcal{B}(t \rightarrow gu)<2.0\times10^{-4}$ and 
$\mathcal{B}(t \rightarrow gc)<3.9\times10^{-3}$.
~

\subsection{Search for heavy $W'$ boson production}
\begin{figure}
\centering
\includegraphics[width=60mm]{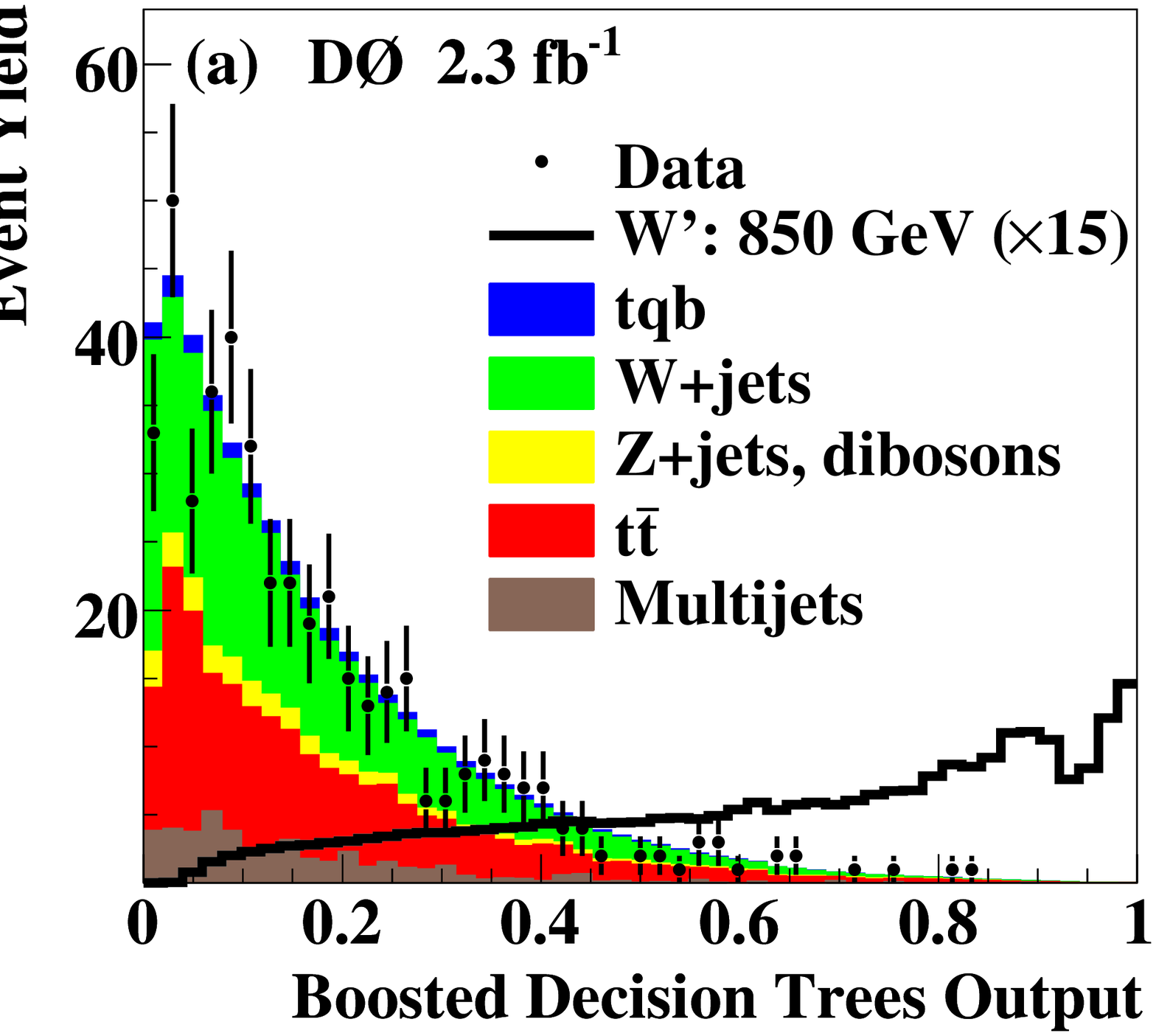}
\includegraphics[width=60mm]{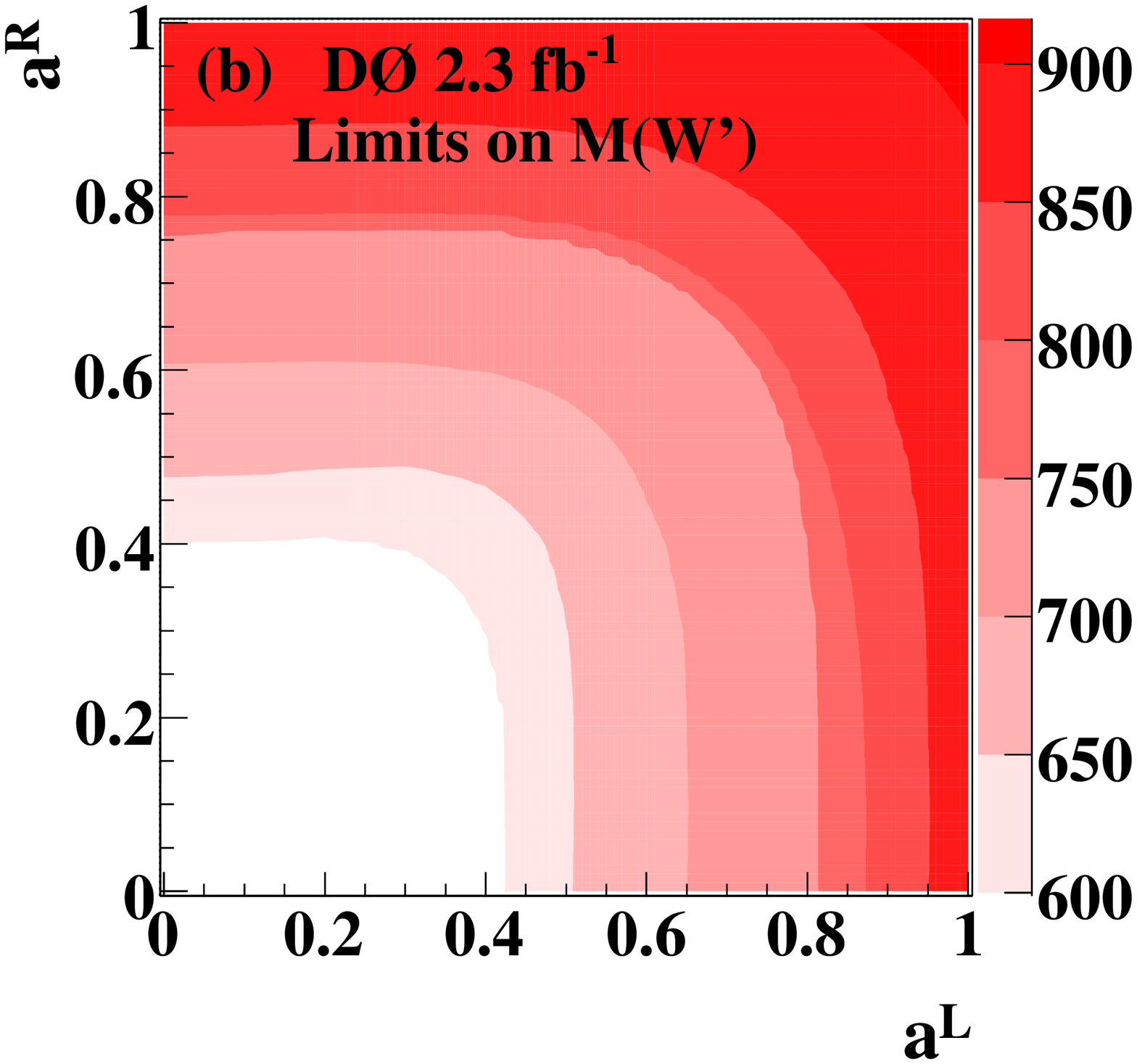}
\caption{(a) Distributions of the BDT discriminant for the $W'$~boson search
for a $W'$~boson mass of 850~GeV, including the expected $W'$~boson signal with 
mixed couplings. (b) Contours of 95\% C.L. lower limits on $M(W^\prime)$ in the 
(aL, aR) plane.
\label{fig:wprime}}
\end{figure}

Models of new physics that introduce additional symmetries predict the existence 
of additional heavy charged bosons, generally called $W'$. 
The D0~collaboration has performed a search for $W'$~boson production in 
2.3~fb$^{-1}$ of data~\cite{Abazov:2011xs}, 
using the same event selection and background modeling
as the single top quark observation analysis~\cite{Abazov:2009ii}. 
This analysis explores the full parameter space of both SM-like left-handed 
couplings ($aL$) and right-handed couplings ($aR$) and a mixture of both for 
the $W'$~boson. 
For each $W'$~boson mass under consideration, a Boosted Decision Tree (BDT)
is trained to separate the $W'$~boson signal from the backgrounds. The BDT
output for one mass is shown in Fig.~\ref{fig:wprime}(a).

The resulting limits on the $W'$~boson mass as a 
function of the two couplings are shown in Fig.~\ref{fig:wprime}(b). 
For SM-like left-handed $W'$~couplings, the limit on the $W'$~boson mass is 
$M(W^\prime)>863$~GeV. For purely right-handed couplings it depends on
whether a right-handed neutrino ($\nu_R$) exists and is lighter than the 
$W'$~boson, the limit is 
$M(W^\prime)>885$~GeV for $M(W^\prime) < m(\nu_R)$ and 
$M(W^\prime)>890$~GeV for $M(W^\prime) > m(\nu_R)$. If both left-handed and 
right-handed couplings are present, the limit is $M(W^\prime)>916$~GeV.
~

\section{Conclusions}
The D0 experiment at the Tevatron has presented updates on the measurement
of single top quark production. Besides a more precise combined single top
quark production cross section measurement, the $t$-channel mode was isolated
for the first time. 
All of the measurements are consistent with the SM expectation.  
Several new physics scenarios have been explored in the single top quark final 
state and limits have been set on on flavor-changing neutral currents and
on new heavy boson $W'$ production.

\bigskip 

\begin{thebibliography}{99} 
\bibitem{Abazov:2009ii}
  V.~M.~Abazov {\it et al.}  (D0 Collaboration),
  Phys.~Rev.~Lett.~{\bf 103}, 092001 (2009).

\bibitem{Aaltonen:2009jj}
  T.~Aaltonen {\it et al.}  (CDF Collaboration),
  Phys.~Rev.~Lett.~{\bf 103}, 092002 (2009).

\bibitem{singletop-vtb-jikia}
G.V.~Jikia and S.R.~Slabospitsky,
Phys.\ Lett.\ B {\bf 295}, 136 (1992).

\bibitem{Chen:2005vr}
  C.R.~Chen, F.~Larios, and C.~P.~Yuan,
  Phys.\ Lett.\ B\ {\bf 631}, 126 (2005).
%
\bibitem{dudko-boos}
E.~Boos, L.~Dudko, and T.~Ohl,
Eur.\ Phys.\ J.\ C {\bf 11}, 473 (1999)

\bibitem{singletop-wtb-heinson}
A.P.~Heinson, A.S.~Belyaev, and E.E. Boos,
Phys.\ Rev.\ D {\bf 56}, 3114 (1997).

\bibitem{d0-singletop-wtb}
V.M.~Abazov {\it et al.} (D0 Collaboration),
Phys.\ Rev.\ Lett.\ {\bf 101}, 221801 (2008).

\bibitem{Abazov:2009ky}
  V.~M.~Abazov {\it et al.}  (D0 Collaboration),
  Phys.\ Rev.\ Lett.\ {\bf 102}, 092002 (2009).

\bibitem{Tait:2000sh}
T.~Tait and C.-P.~Yuan,
Phys. Rev. D {\bf 63}, 014018 (2001).
%

\bibitem{Abazov:2011pt}
  V.~M.~Abazov {\it et al.} (D0 Collaboration),
  {\em submitted to Phys.~Lett.~B},
  [arXiv:1108.3091 [hep-ex]] (2011).

\bibitem{Abazov:2011rz}
  V.~M.~Abazov {\it et al.} (D0 Collaboration),
{\em submitted to Phys.~Rev.~D},  
  [arXiv:1105.2788 [hep-ex]] (2011).

\bibitem{d0-prd-2008}
V.~M.~Abazov {\it et al.} (D0 Collaboration),
Phys.\ Rev.\ D\ {\bf 78}, 012005 (2008).

\bibitem{singletop-xsec-kidonakis}
N.~Kidonakis,
Phys.\ Rev.\ D {\bf 74}, 114012 (2006). 

\bibitem{Abazov:2010qk}
  V.~M.~Abazov {\it et al.} (D0 Collaboration),
  Phys.\ Lett.\  {\bf B693}, 81-87 (2010).
  [arXiv:1006.3575 [hep-ex]].

\bibitem{Abazov:2011xs}
  V.~M.~Abazov {\it et al.}  (D0 Collaboration),
  Phys.\ Lett.\  B {\bf 699}, 145 (2011)
  [arXiv:1101.0806 [hep-ex]].

\end{thebibliography}

\end{document}